\documentclass[aps,prl,twocolumn,groupedaddress]{revtex4}
\usepackage{graphicx}
 \begin{document}
 \title{Flexible lipid bilayers in implicit solvent}
 \author{Grace Brannigan}
\affiliation{Department of Physics and Astronomy, University of California, Santa Barbara, 93106-9530}
\author{Peter F. Phillips}
\author{Frank L.H. Brown}
\affiliation{Department of Chemistry and Biochemistry, University of California, Santa Barbara, 93106-9510}
\begin{abstract}
A minimalist simulation model for lipid bilayers is presented.  Each lipid is represented by a flexible chain of beads in implicit solvent.  The hydrophobic effect is mimicked through an 
intermolecular pair potential localized at the ``water''/hydrocarbon
tail interface. This potential guarantees realistic interfacial tensions for lipids in a bilayer geometry.  Lipids self assemble into 
bilayer structures that display fluidity and elastic properties consistent with experimental model membrane systems. Varying molecular flexibility allows for tuning of elastic moduli and area/molecule over a range of values seen in experimental systems.    
\end{abstract}
\date{\today}
\maketitle


Lipid bilayer biomembranes are of fundamental importance in cellular biology, and model membrane systems are fascinating physical systems in their own right.  Simulation models for
lipid bilayers have been developed over a range of resolutions (from fully atomistic descriptions to continuous elastic sheets) to address the many different length scales relevant to biological function and experimental study.  The ``mesoscopic'' regime ($\sim 1-100 \mathrm{nm}$) is  recognized as particularly relevant to the biophysical understanding of membrane systems \cite{evansreview}.

Several coarse-grained bilayer models have been developed with the mesoscopic regime in mind \cite{maggs, lipowsky, models,takasu,grace}.  Most of these depend upon explicit solvent to enforce bilayer stability.  
Coarse-grained solvent models do not provide detailed  insight into the hydrophobic effect; the models are far too crude.  Rather, such models provide a convenient means to enforce a bilayer stabilizing interfacial tension between solvent and lipid hydrocarbon tails (at considerable computational expense).  Controlling the interfacial tension directly, without 
recourse to explicit solvent, would seem (if possible) a more direct route to the same end.  A 
few solvent-free models \cite{maggs, takasu, grace} for bilayers do exist in the literature, but 
none include internal degrees of freedom for the lipids.  Consequently, these models are unable to predict how molecular structure influences membrane properties, phase behavior or realistic consequences of membrane heterogeneity. Such questions require flexible lipids to achieve even a qualitative level of understanding. \cite{evansreview, oldnagle, evanssaturation}  

This letter presents a solvent-free lipid model that preserves the physics of lipid flexibility and
hydrophobic attraction. The physical properties of the studied membranes closely resemble those of a solvated model with similar lipid resolution.\cite{lipowsky} These results suggest that implicit solvent models may be appropriate for a wide class of problems in membrane biophysics.  In particular, the computational simplicity of the present model makes it very attractive for future studies of heterogeneous bilayers, phase behavior and related phenomena dependent on mesoscale lipid structure.  The elementary approach adopted for mimicking hydrophobic attraction holds promise for extension to lipid models beyond those studied here. 

\begin{figure}
 \begin{center}
\resizebox{3.0in}{1.75in}{\includegraphics{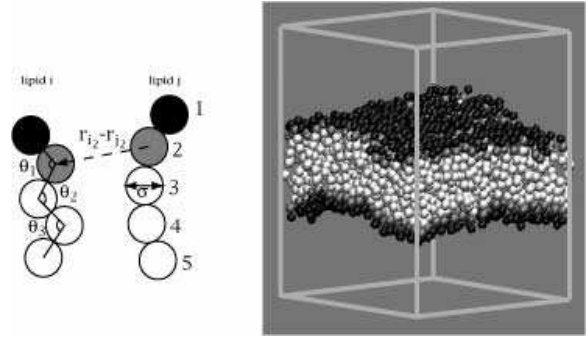}}
\caption{\label{fig:membrane} Left: Definition of parameters used in model. Right: Sample conformation of tensionless membrane with 800 lipid molecules ($c_{bend}=7\epsilon$). Polar head beads are black, interface beads are gray, and hydrophobic tail beads are white.  Simulations presented in this work employ 5 bead lipids exclusively.  Modification to longer lipids with more tail beads is possible and straightforward.}
\end{center}
\end{figure}

Individual lipids are represented as semi-flexible chains of five beads (Fig. \ref{fig:membrane}).  
Bead 1 is identified as the polar head group, bead 2 is associated with the interface between
polar and hydrophobic components and beads 3-5 comprise the hydrophobic tail region. 
Bonded bead-bead distances are constrained to have length $\sigma$ and
bond angles are subject to a bending potential equivalent to that employed 
by Goetz and Lipowsky \cite{lipowsky}:
\begin{equation}
U_{bend}(\theta) = c_{bend}\cos\theta, 
\end{equation}
where $\theta$ is one of three bond angles on the molecule (Fig. \ref{fig:membrane})
and $c_{bend}$ is a positive energetic constant. There is no energetic cost for dihedral
rotations. 

Individual beads interact through a combination of three different 
pair-potentials:
\begin{eqnarray}
U_{core}(r)& =& c_{core}(\sigma/r)^{12}\\
U_{tail}(r)&=& -c_{tail}(\sigma/r)^{6}\\
U_{int}(r)&=& -c_{int}(\sigma/r)^{2}
\end{eqnarray}
where $c_{core}$, $c_{tail}$ and $c_{int}$ are all positive energetic constants. With the exception of intra-molecular bead pairs separated by less than three bonds, the repulsive core interaction acts between all bead pairs and the tail dispersion attraction acts between all tail-interface and tail-tail pairs. The soft interfacial attraction (discussed later) acts between all interface-interface pairs. The potentials are truncated at distances of $2\sigma, 2\sigma$ and $3\sigma$ for the core, tail, and interfacial interactions respectively, and shifted to insure continuity of the potential.  Truncation of the
otherwise long ranged $U_{int}$ is an essential component of the interaction and should not be viewed as an approximation to a true $r^{-2}$ potential.

The results described below were obtained using the values $k_{B}T=0.9\epsilon$, $c_{core} = 0.4\epsilon$, $c_{tail} = 1.0 \epsilon$, and $c_{int} = 3.0\epsilon$ with $c_{bend}$ varied between $5.0-10.0\epsilon$.  Low values of $c_{bend}$ ($<5\epsilon$) resulted in
bilayers with a tendency to form pores and high values of $c_{bend}$ ($>15\epsilon$) gave rise to
bilayers with ordered structure. The reduced units are calibrated by mandating that the simulations are conducted at $300 K$ and that the largest observed area per molecule (at $c_{bend} = 5.0\epsilon$) corresponds to about $0.7$nm$^2$. This results in the unit scale $\epsilon = 2.75$kJ/mol and $\sigma = 0.75$nm.
 
\begin{figure*}
 \begin{center}
{\resizebox{5in}{2.25in}{\includegraphics{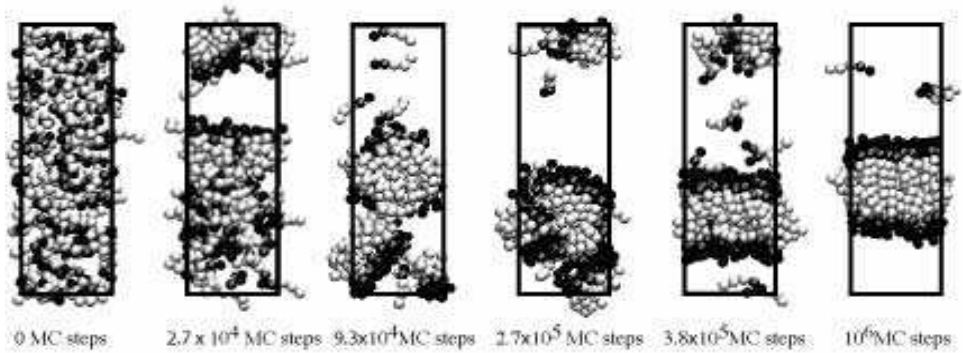}}}
 \caption{\label{fig:assembly} Self-assembly of a bilayer patch of 128 lipids ($c_{bend}=7\epsilon$) in a box with constant dimensions $L_x = L_y = 8.6\sigma, L_z = 25\sigma$ and periodic boundary conditions. The chosen area corresponds to that assumed by a pre-assembled bilayer at zero tension.  Each Monte Carlo ($MC$) step includes (on average) an attempt to translate each bead in the simulation.}
\end{center}
\end{figure*}

Simulations were carried out by Metropolis Monte Carlo using standard moves for short chains
\cite{chains} and an additional move that attempted translation of an 
entire molecule. Stability was verified by bilayer assembly at constant box dimensions from a random configuration of 128 molecules (Fig. \ref{fig:assembly}). All other simulations were conducted in the constant vanishing tension/constant volume ensemble\cite{venturoli, grace}. Crystalline bilayers of 800 molecules were allowed to equilibrate prior to data collection, and fluidity was verified by lateral diffusion.  A density of  $0.07$ lipids/$\sigma^3$ was used throughout, which prevents the membrane from interacting with its periodic images in the $z$ direction. During the course of simulation, lipids occasionally leave the bilayer to explore the box and later reenter the bilayer - i.e. at equilibrium monomers and bilayer lipids exchange ( 0  - 3 \% monomers depending on $c_{bend}$ ).

 
\begin{figure}
 \begin{center}
\resizebox{3in}{2in}{\includegraphics{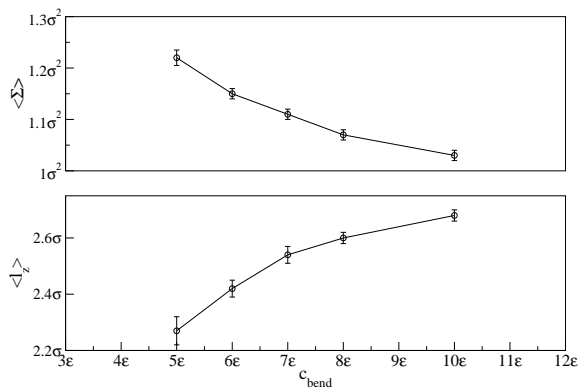}}
\caption{\label{fig:area} Projected area per molecule (top) and leaflet thickness(bottom) as a function of molecular bending coefficient $c_{bend}$  for membranes under zero tension. The volume $\langle \Sigma l_z \rangle$ is insensitive to $c_{bend}$. Lines are to guide the eye.}
\end{center}
\end{figure}

In membranes with $c_{bend} = 5-10\epsilon$, molecular area scales inversely with molecular rigidity, but a simultaneous increase in bilayer thickness preserves membrane volume (Fig. \ref{fig:area}). The thickness of a leaflet is defined by the average $z$ distance between a molecule's interface bead and its final tail bead: $\langle l_z \rangle = \langle (\vec{r}_5-\vec{r}_2)\cdot\hat{z}\rangle$. Unsurprisingly, stiffer chains offer greater resistance to compression in length, resulting in thicker membranes (Fig. \ref{fig:area}).  In model membrane systems, zero tension areas per molecule range from about 0.596 nm$^2$ for DMPC(dimyristoylphosphatidylcholine) to 0.725 nm$^2$ for DOPC(dioleoylphosphatidylcholine).\cite{naglereview}  Using our model, we achieve a $20\%$ range ($~0.57$nm$^2-0.68$nm$^2$) in areas by adjusting the chain stiffness alone. 

A linearly elastic sheet can be described by
\begin{equation}
\label{eq:kaeq}
\tilde{\gamma} = k_A (L^2 - A_0)/{A_0}
\end{equation}
where $\tilde{\gamma}$ is the surface tension, $k_A$ is the compressibility modulus, $L^2$ is the projected area, and $A_0$ is the zero tension area.  Although the simulation algorithm maintains a constant thermodynamic tension $\gamma$, $\tilde{\gamma}$ represents the mechanical surface tension, which is measurable via the Virial stress tensor \cite{frenkel} and fluctuates throughout the simulation (with sufficient averaging, $\langle \tilde{\gamma} \rangle = \gamma$). 
We measure $k_A$ by expressing Eq. \ref{eq:kaeq} as \cite{fellerngt}
\begin{equation}
k_A = \frac{\langle  \tilde{\gamma} L^2 \rangle \langle L^2 \rangle - \langle\tilde{\gamma}\rangle \langle L^4\rangle}{\langle (L^2)^2\rangle - \langle L^2 \rangle^2}, 
\label{eq:llsq}
\end{equation}
and evaluating the averages over the course of our vanishing tension simulations.  
 $k_A$ values range  from $5\pm 4\epsilon/\sigma^2$ for $c_{bend}=5\epsilon$ to $28\pm 9\epsilon/\sigma^2$ for $c_{bend} =10\epsilon$.   Given our unit calibration, these values correspond to 
 40-224 mJ/m$^2$, in good agreement with single component phospholipid bilayers, which typically range from $60-270$ mJ/m$^2$\cite{evanssaturation, sackmann}. In contrast to the results of Refs. \cite{marrink, briels}, we found that $k_A$ measurements for systems with 800 molecules and 128 molecules agreed within error bars, although the smaller system measurements converged far more quickly. 

\begin{figure}
 \begin{center}
\resizebox{3in}{2in}{\includegraphics{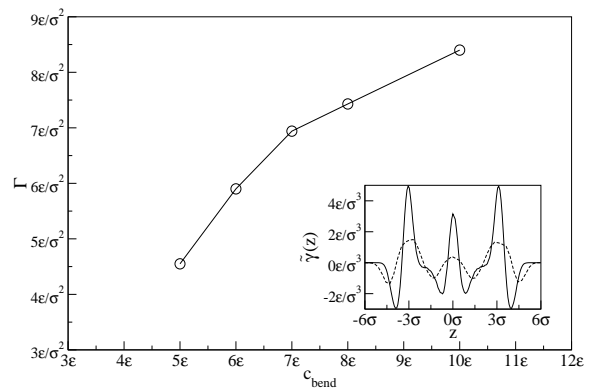}}
\caption{\label{fig:ka}Interfacial tension as defined in Eq. \ref{eq:pieq}. Error bars are smaller than symbol size and line is to guide the eye. Inset: Stress profile for systems with 128 molecules (solid) and 800 molecules (dashed) showing the same pattern of peaks and valleys as observed in a similar solvated model\cite{lipowsky}.  Profiles correspond to systems with $c_{bend}=7\epsilon$.}
\end{center}
\end{figure}

While $k_A$ provides a direct link to experiment, more detailed 
microscopic information is obtained by measuring the stress profile across the bilayer \cite{lipowsky,lindahlstress}.
 Defining the local mechanical tension as a function of displacement $z$ relative to the 
 bilayer center of mass $\tilde{\gamma}(z)=P_n(z) - P_t(z)$ (the difference between normal and tangential pressures), we measure the stress profile for systems with N=128 and N=800 molecules (Fig. \ref{fig:ka} inset).  The profiles agree qualitatively with those obtained from fully atomic models\cite{lindahlstress} and nearly quantitatively with those obtained from solvated membranes also composed of five bead chains \cite{lipowsky}. The peaks of high positive tension
 correspond to the positions of the interface beads, indicating  that these beads are holding
 the bilayer together, against the lateral repulsions of polar heads and third and fourth beads.
 Undulations significantly smooth out the profile, even in moderately sized bilayers with 800 molecules.
 
In our model, a strong attraction between interface beads mimics the hydrophobic effect. Since all ``solvent'' effects of our model are incorporated within $U_{int}$, the effective interfacial tension 
 is given by the interfacial contribution to the Virial tension:
 \begin{equation}
\label{eq:pieq}
{\Gamma} = \frac{1}{2}\sum_{i<j}\left\langle\frac{r_{ij}^2 - 3 z_{ij}^2}{2  L^2r_{ij}}\frac{\partial U_{int}(r_{ij})}{\partial r_{ij}}\right\rangle,
\end{equation}
where the sum is over all distinct pairs of interface beads. The factor of $1/2$ corresponds to the two interfaces present in a bilayer and $\Gamma$
is thus defined in the usual sense of the interfacial tension \cite{israelachvili}. The surface pressure for each leaflet is given by the difference between total and interfacial tensions $\Pi = \Gamma - \langle \tilde{\gamma} \rangle/2$ so that
$\Pi=\Gamma$ in the zero stress state simulated here.   We measure values of $\Gamma = 4-8\epsilon/\sigma^2 (32-65$ mJ/m$^2)$ (Fig.\ref{fig:ka}). Theoretical estimates range from $20 - 50$ mJ/m$^2$\cite{oldnagle, israelachvili, marsh}).

The functional form chosen for $U_{int}$ in our model is empirical, and was identified through trial and error motivated by our
previous experience with rigid lipid models \cite{grace}.  The approximate magnitude of $c_{int}$ given this functional form
is dictated by physical necessity: $c_{int}=3.0\epsilon$ leads to a stable
fluid phase for a variety of $c_{bend}$ values and physically reasonable interfacial tensions.  Connection between $U_{int}$ and the hydrophobic effect is established solely on this basis.
 
 \begin{figure}
 \begin{center}
\resizebox{3in}{2.15in}{\includegraphics{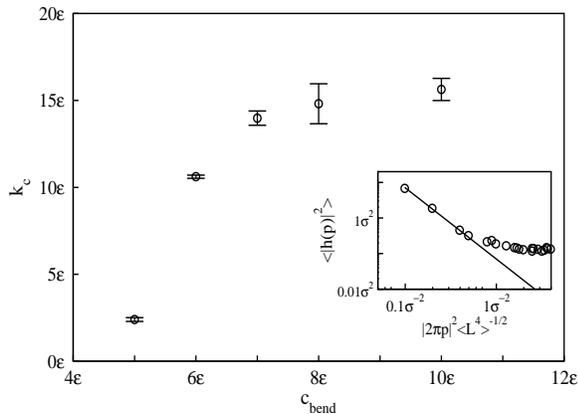}}
\caption{\label{fig:spectrum}Membrane bending rigidity ($k_c$) as a function of molecular rigidity ($c_{bend}$). Data points represent a fit  to the four longest wavelength modes; error bars represent the standard error in the fit. Each data point corresponds to about one week of computation on a 2.3 GHz Athlon CPU.  INSET: Spectrum, for $c_{bend} = 7\epsilon$, where the line is a fit to Eq. \ref{eq:spectrum} with $k_c = 13.9\epsilon$. 
}
\end{center}
\end{figure}
Membranes display obvious long wavelength fluctuations (Fig. \ref{fig:membrane}) that we have quantified via the fluctuation spectrum\cite{maggs, lipowsky}.  In the constant zero tension ensemble, a flexible fluid sheet 
is expected to display undulations consistent with \cite{grace}:
\begin{equation}
\label{eq:spectrum} \langle |h(\vec{p})|^2 \rangle = \frac{k_BT}{k_c}\frac{\langle L^4\rangle}{(2\pi |\vec{p}|)^4}, 
\end{equation}
where $k_c$ is the bending rigidity, $h(x,y)$ is the local height of the membrane midplane,  $h(\vec{p}) = 1/L \int dx dy h(x,y)e^{i 2\pi\vec{p}\cdot \vec{r}/L}$ and the components of $\vec{p}$ are $\big(\pm 1,\pm 2...\pm n/2).$  The inferred rigidities are shown in Fig. \ref{fig:spectrum} with a representative spectrum shown 
in the inset.  Wavelengths shorter than the membrane thickness clearly do not follow Eq. \ref{eq:spectrum} and have been excluded from the fit as discussed extensively in prior work \cite{lipowsky,marrink, thickness}. 
Bending rigidities range from $k_c=2.5-16\epsilon$ (approx.  $1-8*10^{-20}J$). 
The more flexible systems are consistent with experimental measurements of DGDG (digalactosyldiacylglycerol), while the stiffer systems agree well with measurements of DLPC(dilauroylphosphatidylcholine) and DMPC.\cite{seifertinsackmann} Stiffer molecules lead to stiffer membranes; this can be partially attributed to the increase in both compressibility modulus and membrane thickness. It has been suggested that one should generally expect $k_c = k_A d^2/b$, where $d$ is the bilayer thickness and $b$ is a dimensionless constant \cite{evansreview,lipowsky, evanssaturation}.  We measure $b\sim60$ for the model presented here, well within the limits
seen in other simulation models ($b\sim 4$ to $b\sim 100$) \cite{lipowsky,grace,marrink,thickness,briels} and experiment.\cite{evanssaturation}

Mesoscopic models provide a link between atomic level detail and macroscopic
physical properties.  The present description  incorporates a level of realism previously lacking
in implicit solvent models for lipid bilayers and should allow for detailed studies of biophysical
questions where solvated models are computationally prohibitive.  Additionally, the physical
picture afforded by this model is very much in the spirit of analytical theories that seldom 
consider water explicitly.  Modeling at this level of detail should provide a critical link between
experiment, theory and atomistic simulations.  The simulation of inhomogeneous membrane surfaces 
(with multiple lipid species and protein inclusions) is especially promising and is currently
under investigation.

This work was supported in part by the Petroleum Research Fund of the American
Chemical Society (42447-G7) and the National Science Foundation (MCB-0203221,CHE-0321368).
\bibliography{flexible.bib}

  \end{document}